\documentclass[aps,prd,final,twocolumn,letterpaper,twoside]{revtex4}

\usepackage{amsmath, amssymb, amsfonts}
\usepackage{graphicx}
\usepackage{microtype}
\usepackage[tight]{subfigure}

\usepackage[normalem]{ulem}
\usepackage{enumerate}

\newcommand{\bra}[1]{\langle #1|}
\newcommand{\ket}[1]{|#1\rangle}
\newcommand{\braket}[2]{\langle#1 |  #2\rangle}

\newcommand{\T}{\mathsf{T}}

\begin{document}
\title{A Brief Introduction to Band Structure in Three Dimensions}
\author{P.A.~Iannucci}
\affiliation{ Department of EECS,
 77 Massachusetts Ave.,
Cambridge, MA 02139-4307}
\date{\today}

\begin{abstract}
\noindent

Without our ability to model and manipulate the band structure of semiconducting
materials, the modern digital computer would be impractically large, hot, and
expensive.  In 8.06, we studied the effect of spatially periodic potentials on
the spectrum of a charged particle in one dimension.  We would like to
understand how to extend these methods to model actual crystalline materials.
Along the way, we will explore the construction of periodic potentials in three
dimensions, and we use this framework to relate the single-particle Hamiltonian
to the potential contribution from each atom.  We then construct a crude model
system analogous to the semiconductor silicon, and demonstrate the appearance of
level splitting and band gaps as the strength of the potential is varied, in
accordance with our intuition from the one-dimensional case.  We discuss
refinements of the model to include many-particle effects, and finally we show
how a careful choice of the potential function leads to good agreement with the
correct band diagram for silicon.
\end{abstract}

\maketitle

\section{Introduction}
\noindent
Many world-changing results from solid state physics, like the development of
the silicon transistor, take advantage of the band structure of crystalline
materials to give engineers detailed control over material behavior.  Because
crystals are ultimately many-particle systems with many degrees of freedom, the
models that are used to describe these materials make a wide range of
approximations in order to obtain mathematically and computationally tractable
results about band structure.  The most detailed models can predict not only the
number and spacing of bands and the density of available states in the material,
quantities useful for understanding the behavior of electrons near the Fermi
surface at nonzero temperature, but also the effects of mechanical strain,
defects, and impurities on the crystal's electronic properties.  Semiconductor
engineers fit these models to empirical data to obtain very nuanced control over
the behavior of devices.

In 8.06, we explored the behavior of single electrons in one-dimensional periodic
potentials.  This toy model allowed us to invoke several of the most
important techniques for learning about band structure, including the
nearly-free electron model, which explained how tiny interactions between
otherwise free valence electrons can open up band gaps, and the tight-binding
model, which showed that even closely-held core electrons undergo level
splitting to form narrow bands.

In this work, we will show how these techniques generalize from one dimension to
three dimensions, demonstrating the appearance of level splitting and band gaps
as a periodic potential is gradually ``switched on''.  We will also show how an
approximation to the correct band structure of the semiconductor silicon can be
obtained within this model by adjusting the potential function.

In order to apply the nearly-free electron model, we will need a
mathematical framework for the three-dimensional lattice.  We can then take a
periodic potential on this lattice over to the Fourier domain, recovering
the reciprocal lattice.  This will tell us how the Brillouin zone generalizes
to three dimensions.  After reviewing Bloch's theorem, we will put all of these
pieces together to find and solve the Hamiltonian for a particle moving in a
three-dimensional periodic potential, and consider briefly how this approach can be
extended to the many-particle case.

\section{Lattice Formalism}
\subsection{Bravais Lattice}
\noindent
Consider the potential energy of an electron located at a point $\vec x$ within an
infinite grid of positively-charged ions, all carrying the same charge.  We can
write this as
\begin{equation}
V(\vec x) = \sum_i U(\vec x - \vec x_i)
\label{potential}
\end{equation}
where $x_i$ is the position of the $i^{\mathrm{th}}$ atom, and $U(\vec x)$ is
the potential due to a single ion at the origin.  In one dimension, we write
the positions of the atoms as
\begin{equation}
\vec x_n = n a \hat x = n\vec a_1
\end {equation}
for integer $n$, where $a$ is the lattice spacing, and we have defined $\vec a_1
\equiv a \hat x$.  The region of 1-D space bounded by $(0, a)$ is
a unit cell.

Notice that there is no particular reason to require that $U(\vec x)$ be the
potential due to a single ion.  One could construct a 1-D crystal where small,
identical clusters of atoms form a repeating pattern with period $a$, with
each cluster containing several atoms spaced by less than $a$.

\begin{figure*}[ht]
\includegraphics[scale=0.3,clip=true,angle=0]{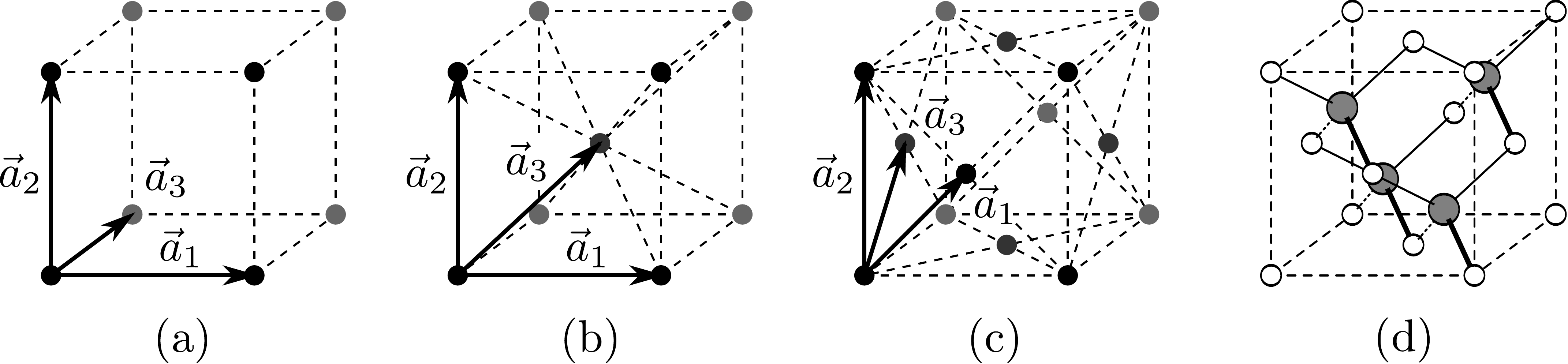}
\caption{Cubic Bravais lattice system: (a) simple cubic, (b) body-centered
cubic, (c) face-centered cubic.  The vectors $\vec a_{1,2,3}$ represent directions of
periodicity.  The unit cells contain 1, 2, and 4 points, respectively.  (d)
Diamond cubic variation, with additional atoms highlighted.  The lines
illustrate the tetrahedral symmetry about each atom.  Atoms in (d) connected by a thick
line are associated with the same lattice point.}
\label{fig1}
\end{figure*}

In three dimensions, the situation is more interesting, since we have some
freedom to define the directions along which the crystal is periodic (they don't
necessarily have to be perpendicular).  Even for oblique directions of
periodicity, we can write
\begin{equation}
\vec x_{n,m,\ell} = n\vec a_1 + m\vec a_2 + \ell\vec a_3
\end {equation}
to give the positions of a regular grid of points.  Families of points
describable in these terms are called Bravais lattices.  Common Bravais lattices
include the cubic lattice system (simple, body-centered, and face-centered) and
the hexagonal lattice.  The cubic lattice system is shown in
Figure~\ref{fig1}(a), (b), and (c), with circles indicating the locations of
atoms in the simple, body-, and face-centered cubic configurations,
      respectively.  You should take a moment to convince yourself
that shifting the origin by one step along any of the indicated directions of
periodicity $\vec a_{1,2,3}$ leaves the lattice unchanged.

As in the 1-D case, it should be stressed that the potential $U(\vec x)$ can
describe a cluster containing more than one ion.  For instance, diamond,
silicon, germanium, and the allotrope $\alpha$-tin adopt a crystal structure that positions
two atoms at each point of a face-centered cubic (FCC) lattice.  This structure,
known as ``diamond cubic'', is shown in Figure~\ref{fig1}(d).

\subsection{Bloch's Theorem}
\noindent
If a Hamiltonian commutes with a set of unitary, mutually commuting
translation operators $\hat{T}_{\vec a}$, then the eigenstates of $\mathcal{H}$
can be chosen to be (simultaneously) eigenstates of all the $\hat{T}_{\vec a}$.
This follows because hermitian and unitary operators are normal operators, and a
set of normal operators commutes if and only if it is simultaneously
diagonalizable (an extention of our 8.05 result for commuting hermitian operators).
Now any two translation operators of the form $\hat{T}_{\vec a} = \exp(-i\hat{\vec
p}\cdot\vec a/\hbar)$ commute, since the individual components of $\hat{\vec p}$
commute.  Furthermore, since the translation operators are unitary, their
eigenvalues have unit magnitude.

Suppose that $\mathcal{H}$ can be written as the sum of a kinetic term
$\hat{\vec p}\,^2/2m$ and a potential term which satisfies $V(\vec x) = V(\vec x
+ \vec a)$ for some set of $\vec a$.  Then $[\mathcal{H},\hat{T}_{\vec a}]=0$,
and we can find an eigenbasis for $\mathcal{H}$ which satisfies
\begin{equation}
\hat{T}_{\vec a}|\psi\rangle = e^{i\theta_{\vec a}} |\psi\rangle \label{snoopy}
\end{equation}
for $\vec a$ in our set.

We know that composing two translation operators
$\hat{T}_{\vec a}$ and $\hat{T}_{\vec b}$ gives $\hat{T}_{\vec a + \vec b}$.
To maintain consistency with \eqref{snoopy}, $\theta_{\vec a}$ must be linear in
$\vec a$.  We will take
\begin{equation}
\theta_{\vec a} \equiv -\kappa \cdot \vec a
\end{equation}
This gives
\begin{align*}
\langle \vec x|\hat{T}_{\vec a}|\psi\rangle & = e^{-i\kappa \cdot \vec a} \langle\vec
x|\psi\rangle \\
\psi(\vec x - \vec a) & = e^{-i\kappa \cdot \vec a}\psi(\vec x)
\end{align*}
So we see that the eigenstates of $\mathcal{H}$ can be written in the spatial
basis as
\begin{equation}
\psi(\vec x) = e^{i\vec \kappa\cdot \vec x} u(\vec x)
\end{equation}
where $u(\vec x) = u(\vec x + \vec a)$ for all $\vec a$ in our set.

\subsection{Reciprocal Lattice}
\noindent
When we apply Bloch's theorem to the potential
\eqref{potential}, we find that the wavefunction can be written in
terms of a function $u(\vec x)$ which has the same periodicity as the lattice:
\begin{equation}
u(\vec x) = u(\vec x + n \vec a_1 + m \vec a_2 + \ell \vec a_3) \;\forall\;
n,m,\ell \in \mathbb{Z}
\end{equation}
It is productive to ask whether the Fourier transform of such a function (that is,
its representation in the momentum basis) has any helpful properties.  Starting with 
\begin{equation*}
u(\vec x)=u(\vec x+\vec a)
\end{equation*} for some $\vec a$, and writing
each side in terms of the inverse transform of $\widetilde{u}(\vec k)$, we find
\begin{equation*}
\int d^3\vec k e^{i\vec k\cdot\vec x}\widetilde{u}(\vec k) = \int d^3\vec k
e^{i\vec k\cdot\vec x} e^{i\vec a\cdot\vec k} \widetilde{u}(\vec k)
\end{equation*}
Collecting terms and taking the inverse transform gives
\begin{equation*}
0 = (1 - e^{i\vec a\cdot\vec k}) \widetilde{u}(\vec k)
\end{equation*}
from which it follows that
\begin{equation*}
\forall \vec k, \text{ either } \vec a\cdot\vec k = 2\pi N 
\text{ for some $N\in\mathbb{Z}$, or } \widetilde{u}(\vec k) = 0.
\end{equation*}
This result shows that if a function is periodic in $\vec a$, then its
transform must be zero everywhere except where $\vec a \cdot \vec k$ is a
multiple of $2 \pi$.  This can be summarized by saying that periodic functions
have discrete transforms.

In three dimensions, we have three linearly independent directions of
periodicity, and hence three discreteness constraints on the $\vec k$ where
$\widetilde{u}(\vec k)$ is allowed to be nonzero.  Writing $\vec a\cdot\vec
k$ as $a^\T \vec k$, we can combine the three constraints into one matrix expression:
\begin{align*}
\left[\begin{matrix}a_1^\T \\ a_2^\T \\ a_3^\T\end{matrix}\right] \vec k & =
2\pi\left[\begin{matrix}N\\M\\L\end{matrix}\right] \\
\rightarrow \vec k & = 2\pi\left[\begin{matrix}a_1^\T \\ a_2^\T \\ a_3^\T\end{matrix}\right]^{-1}
\left[\begin{matrix}N\\M\\L\end{matrix}\right] \label{recip}
\end{align*}
If we define $\vec g_{1,2,3}$ to be $2\pi$ times the three columns of the inverse
matrix, then we have an expression for a new Bravais lattice:
\begin{equation}
\vec k = N \vec g_1 + M \vec g_2 + L \vec g_3
\end{equation}
Each $\vec k$ vector allowed by this lattice corresponds to a complex
exponential $e^{i\vec k\cdot \vec x}$ in real space which is periodic in $\vec
a_{1,2,3}$.  This structure is called the reciprocal (or dual) lattice, and
the space it occupies has many names: it is alternatively called reciprocal space, dual space,
$\vec k$-space, transform space, or (loosely) momentum space.  From now on, we will refer to the
original Bravais lattice as the real space lattice.  Curiously, if
the real space lattice is face-centered cubic, the reciprocal lattice turns out
to be body-centered cubic, and vice versa.

The preceding argument established that the Fourier transform of a periodic
function is nonzero only on the reciprocal lattice.  If we write the potential
\eqref{potential} as a convolution product,
\begin{equation*}
V(\vec x) = U(\vec x) * \sum_{n,m,\ell} \delta(\vec x - n\vec a_1 - m\vec a_2 - \ell\vec a_3)
\end{equation*}
we can take the Fourier transform of the two factors separately to obtain
\begin{equation*}
\widetilde{V}(\vec k) = \widetilde{U}(\vec k) \sum_{n,m,\ell} e^{-i\vec k\cdot(n\vec a_1 + m\vec a_2 + \ell\vec a_3)}
\end{equation*}
When we carry out the summation, we get a new sum over the reciprocal lattice,
which we can again write in matrix form:
\begin{align*}
(2\pi)^3\!\!\!\sum_{N,M,L} \!\! & \delta(\vec k\cdot\vec a_1 \!-\! 2\pi N) \delta(\vec k\cdot\vec a_2 \!-\! 2\pi M) \delta(\vec k\cdot\vec a_3 \!-\!  2\pi L) \\
& = (2\pi)^3\sum_{N,M,L} \delta\left(\left[\begin{matrix}a_1^\T\\a_2^\T\\a_3^\T\end{matrix}\right]\vec k - 2\pi \left[\begin{matrix}N\\M\\L\end{matrix}\right]\right)
\end{align*}
Finally, we can show by $u$-substitution that this is just
\begin{align}
\frac{(2\pi)^3}{\mathrm{det}\left|a_1 a_2 a_3\right|}\sum_G \delta(\vec k - \vec G) \notag
\end{align}
where $\sum_G$ is over the reciprocal lattice.  So
\begin{align}
\widetilde{V}(\vec k) & = \widetilde{U}(\vec k) \frac{(2\pi)^3}{\mathrm{det}\left|a_1 a_2 a_3\right|}\sum_G \delta(\vec k - \vec G) \notag \\
& = \frac{(2\pi)^3}{\Omega}\sum_G U_G \, \delta(\vec k - \vec G) \label{fourier}
\end{align}
where we have defined $U_G \equiv \widetilde{U}(\vec G)$ and $\Omega \equiv
\mathrm{det}\left|a_1 a_2 a_3\right|$.

\subsection{First Brillioun Zone}
\begin{figure}[ht]
\includegraphics[scale=0.3,clip=true,angle=0]{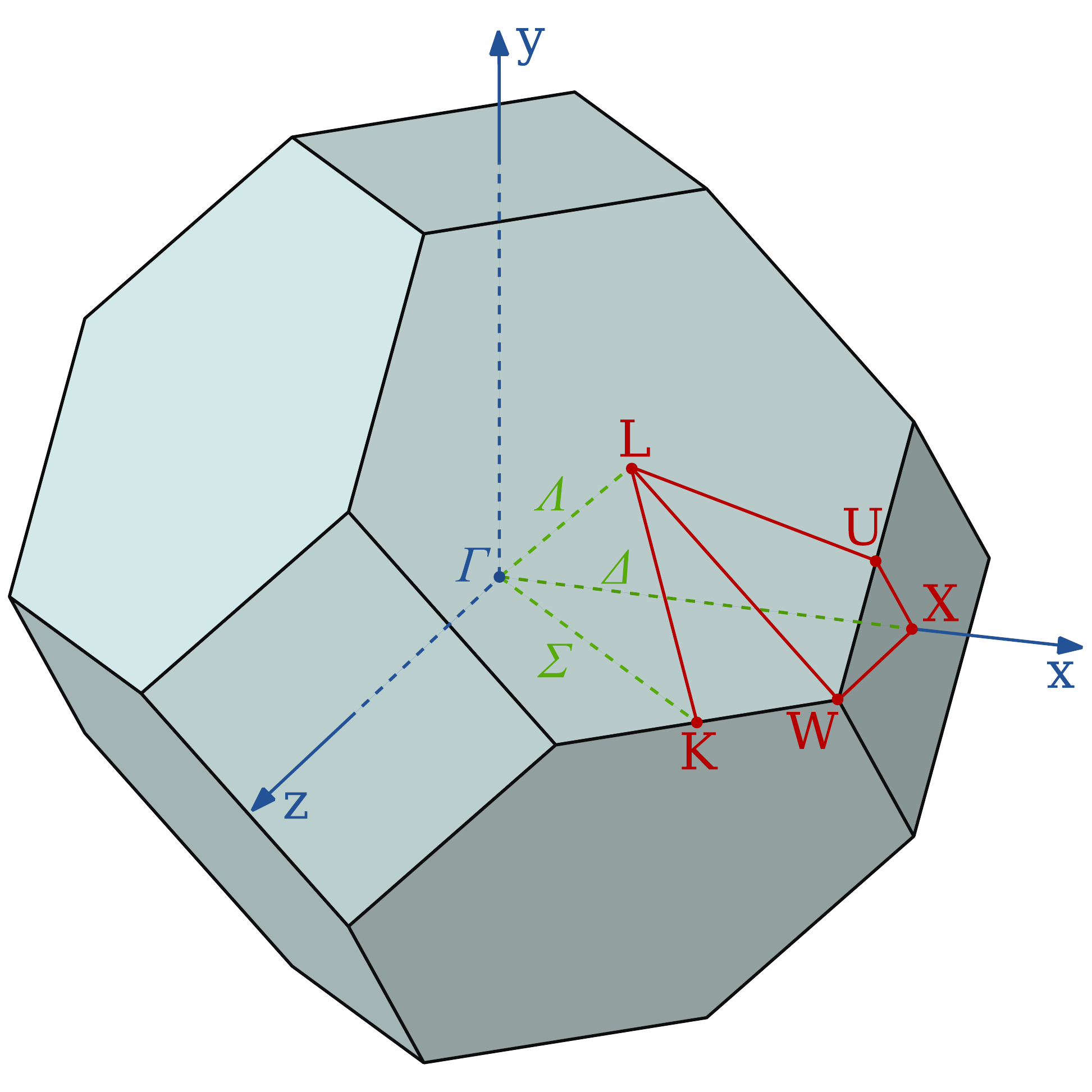}
\caption{First Brillouin zone for a face-centered cubic real space lattice.
Represents the volume in $\vec k$-space closer to $\vec 0$ than to any other
reciprocal lattice vector, and includes all the Bloch wave vectors $\vec \kappa$
necessary to describe the band structure of the crystal.}
\label{bz}
\end{figure}

\noindent
Our study of Bloch's theorem led us to wavefunctions of the
form $\psi(\vec x) = e^{i\vec \kappa\cdot \vec x} u(\vec x)$, with $u(\vec x)$ 
periodic in real space.  But we just observed that every $e^{i\vec
G\cdot \vec x}$ with $\vec G$ on the reciprocal lattice is periodic in real
space.  Thus if we write $\psi'(\vec x) = e^{i(\vec \kappa - \vec G)\cdot\vec x}
(e^{i\vec G\cdot\vec x} u(\vec x)) = e^{i\vec\kappa'\cdot\vec x} u'(\vec x)$, we
obtain an equally valid expression for the exact same wavefunction.  In this
sense, $\vec \kappa$ is not really a free parameter of the system, but can be
restricted without loss of generality to lie in a small region of the reciprocal
space.  Conventionally, we choose this region to contain all vectors $\vec
\kappa$ which
are closer to $\vec G=0$ than to any other reciprocal lattice point.  If we 
start with a $\vec \kappa$ closer to some other reciprocal lattice point $\vec G \neq
0$, we can subtract $\vec G$ from it to get an equivalent point closest to $0$.
The region of space closer to $0$ than to any other reciprocal lattice point is
called the first Brillouin zone.  In one dimension it was given by the
interval $(-\frac\pi a, \frac\pi a)$, since the boundaries between regions
in one dimension are simply points.  In three dimensions, these
boundaries are planes, and the first Brillouin zone has a rather beautiful shape
formed by the intersections of boundary planes.  The first Brillouin zone for an 
FCC real space lattice is shown in Figure~\ref{bz}.  This polyhedron bounds the 
region of reciprocal space where we will focus our attention.

The job of a band structure diagram is to summarize all the interesting things
going on in $\vec k$-space.  Since it is
difficult to capture all the features of a volume on a flat piece of
paper, these diagrams typically take a little tour around the first Brillouin
zone, following a (somewhat) standardized trajectory.
You will notice that a number of particularly symmetric
points in the figure have been named with capital Roman
and Greek letters.  These letters will appear along the horizontal axis of band
structure diagrams, and it is understood that motion along the horizontal axis
corresponds to progress along a trajectory between these points of symmetry in
$\vec k$-space.  For instance, the segment of the graph labeled
\mbox{$\Gamma$--$X$}
corresponds to the straight line in $\vec k$-space from $\vec k=0$ to $\vec
k=(2\pi/a)\hat x$.

\section{Nearly-Free Electron Model}
\noindent
We would like to write down and solve the Hamiltonian for a single
particle moving in a periodic potential.  While it is not true in general that
the dynamics of a many-electron system can be described in terms of one set
of single-particle states, single-particle wavefunctions are nevertheless a
useful tool for approximating the correct many-particle system.  

Working in the plane wave basis, we will find an expression for the matrix
elements of the Bloch Hamiltonian in terms of the Fourier transform of the
potential.  We neglect electron-electron interactions.  Later on,
we will show how to interpret this approximation as an application of
the variational principle.
\subsection{Basis}
\noindent
Our orthogonal basis and normalization convention are
\begin{align}
\braket{\vec x}{\vec x'} & = \delta(\vec x - \vec x') \notag \\
\ket{\vec k} & = \int d^3\vec x e^{i\vec k\cdot\vec x}\ket{\vec x} \\
\braket{\vec k}{\vec k'} & = \int d^3\vec x\int d^3\vec x' \langle\vec x|e^{-i\vec k\cdot\vec x}e^{i\vec k'\cdot\vec x'}|\vec x'\rangle \notag \\
& = (2\pi)^3\delta(\vec k - \vec k')
\end{align}
\subsection{Hamiltonian}
\noindent
We begin with the Hamiltonian for a particle in an arbitrary potential.
\begin{equation}
\mathcal{H} = \frac{\hat{\vec p}\,^2}{2m} + V(\hat{\vec x})
\end{equation}
Using the fact that $\left[\mathcal{H}, \hat{T}_{\vec a}\right] = 0$ for any
translation $\vec a$ in the real space lattice and applying Bloch's theorem, we can 
write the eigenstates $\ket\psi$ as
\begin{equation}
\ket\psi = e^{i\vec \kappa\cdot \hat{\vec x}} \ket{u_{\vec \kappa}}
\end{equation}
where $u_{\vec \kappa}(\vec x)$ has the same periodicity as the real space lattice,
and the Schr\"odinger equation is written as
\begin{align}
\mathcal{H}_{\mathrm{eff}} \ket{u_{\vec \kappa}} & = E \ket{u_{\vec \kappa}} \\
\mathcal{H}_{\mathrm{eff}} & = \frac{1}{2m}(\hat{\vec p} + \hbar\vec \kappa)^2 +
V(\hat{\vec x}) \label{akin}
\end{align}
Using the plane wave basis for $\ket{u_{\vec \kappa}}$, we find matrix elements
\begin{equation}
\bra{\vec k}\mathcal{H}_{\mathrm{eff}} \ket{\vec k'} = \frac{\hbar^2}{2m}(\vec k
+ \vec \kappa)^2\braket{\vec k}{\vec k'} + \bra{\vec k}V(\hat{\vec x})\ket{\vec k'}
\end{equation}
We can write $V(\hat{\vec x})$ in terms of $\widetilde{V}(\vec x)$ with the help of
\eqref{fourier}:
\begin{align*}
V(\hat{\vec x}) & = \frac1{(2\pi)^3}\int d^3\vec kV(\vec k)e^{i\vec k\cdot\hat{\vec x}} \\
& = \frac1\Omega\sum_G U_G \int d^3\vec k \delta(\vec k - \vec G)e^{i\vec k\cdot\hat{\vec x}}
\end{align*}
Using the sifting property of the delta function,
\begin{equation*}
V(\hat{\vec x}) = \frac1\Omega\sum_G U_G e^{i\vec G\cdot\hat{\vec x}}
\end{equation*}
We can now find matrix elements for our potential using the definition of
$|\vec k\rangle$:
\begin{align*}
\bra{\vec k}V(\hat{\vec x})\ket{\vec k'} & = \frac1\Omega\sum_G U_G \!\!\int \!d^3\vec x \! \int d^3\vec x' \bra{\vec x}e^{-i\vec k\cdot\vec x}e^{i\vec G\cdot\hat{\vec x}}e^{i\vec k'\cdot\vec x'}\ket{\vec x'} \\
& = \frac1\Omega\sum_G U_G \!\!\int \!d^3\vec x e^{i\vec x\cdot(\vec G-\vec k + k')}
\end{align*}
Taking the integral, we have
\begin{equation}
\bra{\vec k}V(\hat{\vec x})\ket{\vec k'} = \frac1\Omega(2\pi)^3\sum_G U_G
\delta(\vec G-\vec k + k') \label{grumpy}
\end{equation}
Computations in a continuous basis like $|\vec k\rangle$ are problematic because
they are difficult to approximate numerically.  We would much rather work in a
vector space with a discrete basis, so that our linear operators become ordinary
matrices.  Fortunately, the structure of \eqref{grumpy} permits us to
divide the continuous basis into a set of discrete subspaces, with the assurance
that there will be no coupling between subspaces.  In matrix terminology, our
Hamiltonian is block diagonal.  Using this fact, we can easily restrict the
action of the Hamiltonian to a given subspace and solve numerically.

From \eqref{grumpy}, we can see that the Hamiltonian \eqref{akin} couples states
$\ket{\vec k}$ and $\ket{\vec k'}$ if they differ by a reciprocal lattice vector.  
Otherwise, the off-diagonal matrix elements vanish.  Thus, to every $\vec k$ 
in the first Brillouin zone there corresponds a family of states $\ket{\vec k +
\vec G}$ which are (potentially) coupled by the potential.  Without losing any
physics, we can restrict the Hamiltonian to act in one of these discrete
orthonormal subspaces by constructing (and normalizing) matrix elements
\begin{align}
\frac{\langle \vec k + \vec G | V(\hat{\vec x}) | \vec k + \vec G' \rangle}{\sqrt{\langle \vec k + \vec G'|\vec k + \vec G'\rangle\langle
    \vec k + \vec G|\vec k + \vec G\rangle}} & = \frac{\frac1\Omega (2\pi)^3U_{G-G'} \delta(0)}{\sqrt{(2\pi)^3\delta(0)(2\pi)^3\delta(0)}} \notag \\
& = \frac1\Omega U_{G-G'} \label{pot}
\end{align}
We can build one of these matrices for each $\vec k$ we are interested in,
add $\hbar^2(\vec k + \vec G)^2/2m$ to the diagonal, and we will have the full
Hamiltonian for the corresponding subspace.

Now that we have matrix elements, why don't we apply perturbation theory?
If the potential is small compared to
$\hbar^2(\vec k + \vec \kappa)^2/2m$, we can view it as a perturbation on the
Hamiltonian for a free particle.  In order to obtain corrections to the free
particle energies and wavefunctions, we must first look for the ``good''
states, that is, the basis in which the perturbation is diagonal within every
degenerate subspace of the unperturbed Hamiltonian.  We know that the degenerate
subspaces of the unperturbed (free) Hamiltonian 
correspond to spheres of constant $|\vec k|$ in momentum space.  The question is
whether the matrix element of $V(\hat{\vec x})$ between two states on
the same sphere is nonzero.  Since 
a sphere has chords of every length in $(0,2r]$, we can see by \eqref{pot} that
any vector $\vec G$ in the reciprocal lattice which is short enough to fit
inside the sphere (that is, with $0 < |\vec G| \le 2|\vec k|$) couples together 
degenerate states.  Certain
special potentials, for instance those which are periodic in $x$ but {\em
constant} in $y$ and $z$, permit us to choose a sine and cosine basis that
diagonalizes $V(\hat{\vec x})$.  For more general potentials, finding a good
basis is bothersome. At this 
point, the easiest thing to do is to throw the problem at a computer.  We 
have to make a few more decisions -- most importantly, what our potential function 
$U(\vec x)$ is, and how many states $\ket{\vec k + \vec G}$ to include in the 
truncated (finite) matrix form of $\mathcal{H}_\text{eff}$ -- but given these, finding the 
single-particle energies and wavefunctions is a simple matter of computing each
matrix element numerically using \eqref{pot}, and then invoking a subroutine to
find the eigenvalues and eigenvectors of this matrix.  Fast and accurate
subroutines to perform this function are present in every standard linear algebra
system.  In the popular LAPACK library (Linear Algebra PACKage), for instance,
there is a subroutine called {\tt{zheevd}} that does what we want.  To generate
a band structure diagram, we will need to compute and plot these eigenvalues for
each $\vec k$ we encounter on our tour of the first Brillouin zone.

\section{Model System}
Using the framework we have developed, we can now choose a particular system and
plot some band diagrams to verify our work and improve our intuition into the
phenomenology of band structure.
\subsection{Potential}
\noindent
As it turns out, a great deal of research
\cite{cohen1976,PhysRevB.54.11169,tafipolsky:174102,PhysRevB.26.4199,Pickett1989115} has gone into the
choice of the potential function, since much of the actual many-particle physics
can be incorporated by a suitable choice of $U(\vec x)$.  For instance, once we
obtain wavefunctions for a crystal filled with non-interacting electrons, we
can calculate the effect of e$^-$--e$^-$ repulsion (classical) and correlation
(the quantum ``exchange force'').  Treating these as an adjustment to the
potential, the system can re-solved to obtain a set of wavefunctions that are
sensitive to many-particle effects.  These can in turn be used to make a better
estimate of repulsion and correlation effects.  The iterations stop when the
wavefunction-dependent potential and the potential-dependent wavefunctions are
mutually consistent.  This is the Hartree-Fock method.  The approximation here
lies in the assumption that the many-particle ground state can be written as an
antisymmetrized product of single-particle wavefunctions.  In general, it is a
linear combination of many such products.  This makes Hartree-Fock a variational
technique, since we are effectively minimizing the energy of a trial
multi-particle wavefunction.  The minimization is over all possible sets of
single-particle states.  Since the expectation value of the Hamiltonian on any
state is an upper bound on the ground state, the Hartree-Fock method gives an
upper bound on the energy of the crystal system.

For simplicity, we will use a Coulomb potential of the form
\begin{equation}
U(\vec x) = - \frac{Z_\text{eff}q^2}{|\vec r|}
\end{equation}
Since our goal is to model the diamond-cubic silicon crystal, we will have to 
add another Coulombic potential at an offset to represent the two-ions-per-lattice-site 
structure.  Even this gross simplification (we haven't considered \mbox{e$^-$--e$^-$} 
interactions, spin, etc.) is sufficient to capture some of the flavor of the 
correct band diagram.  In any event, we'll need the Fourier transform of this potential, 
so we'd better go ahead and compute it.

As it turns out, the Fourier integral for the function $-Z_\text{eff}q^2/r$ doesn't converge.  The 
transform of the related ``Yukawa'' potential $-Z_\text{eff}q^2e^{-mr}/r$ is
\begin{equation}
\widetilde{U}(\vec k) = -Z_\text{eff}q^2\frac{4\pi}{k^2 + m^2}
\end{equation}
In the limit as $m \rightarrow 0$, we recover the transform of the Coulomb potential 
everywhere except at $\vec k = 0$.
\begin{equation}
\widetilde{U}(\vec k) = -Z_\text{eff}q^2\frac{4\pi}{k^2}
\end{equation}
Since $U_0$ only appears on the diagonal of the Hamiltonian, ignoring the 
divergence at $\vec k=0$ corresponds to neglecting a constant energy shift.

\subsection{Results}
\begin{figure*}[ht]
\vspace{-5pt}
\subfigure{\includegraphics[scale=0.4,clip=true,angle=0]{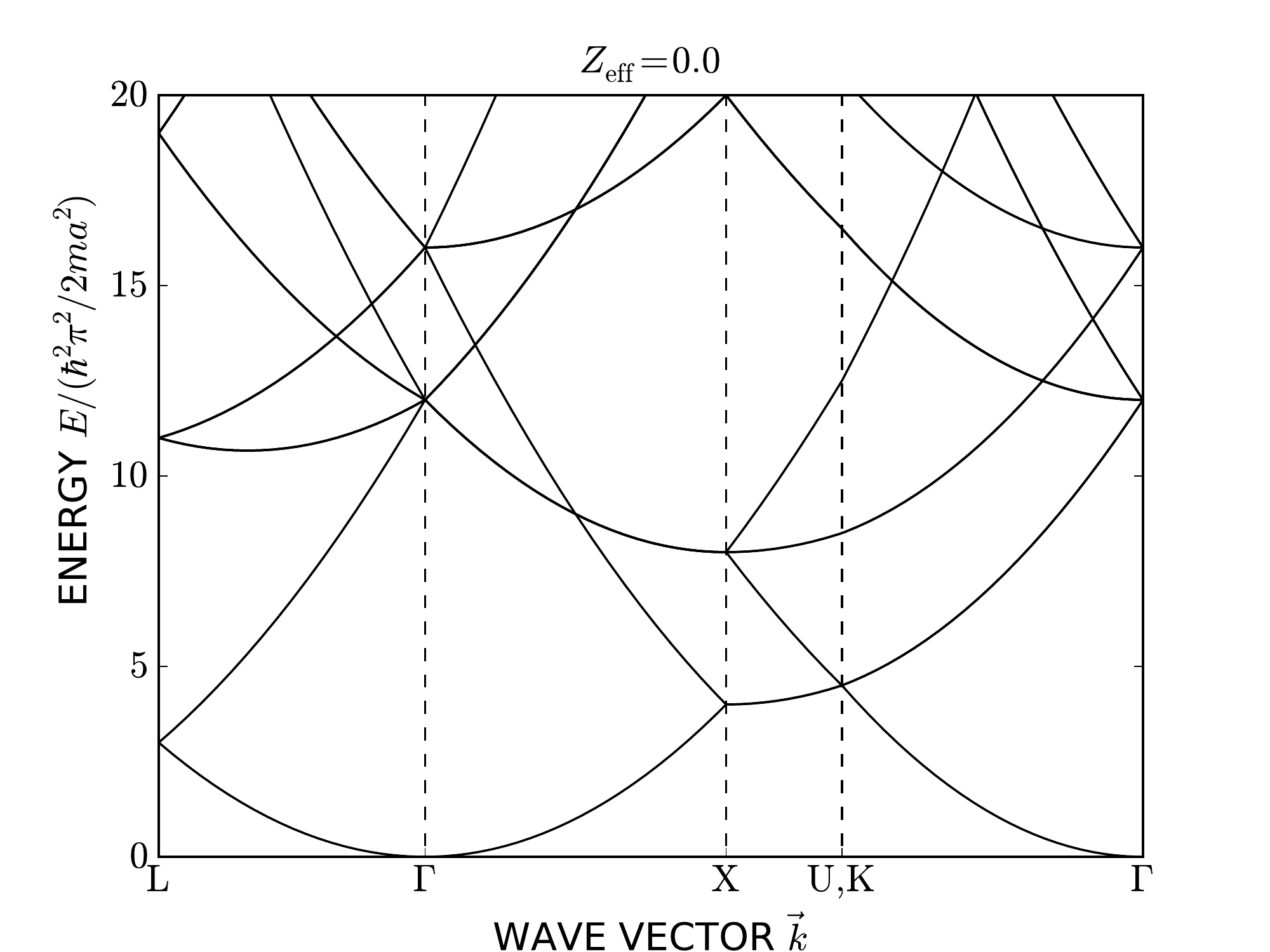}}%
\subfigure{\includegraphics[scale=0.4,clip=true,angle=0]{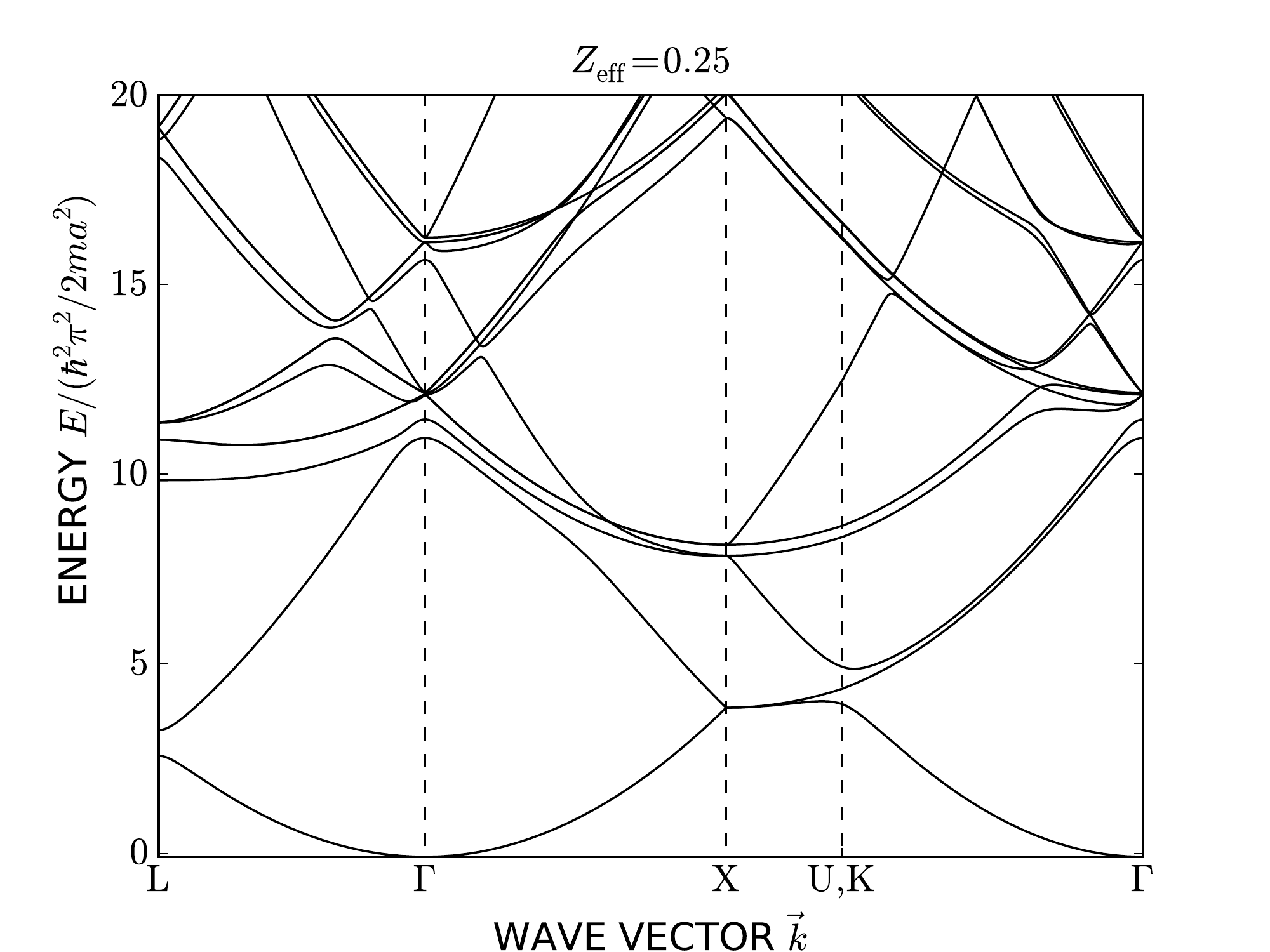}}\\
\vspace{-5pt}
\subfigure{\includegraphics[scale=0.4,clip=true,angle=0]{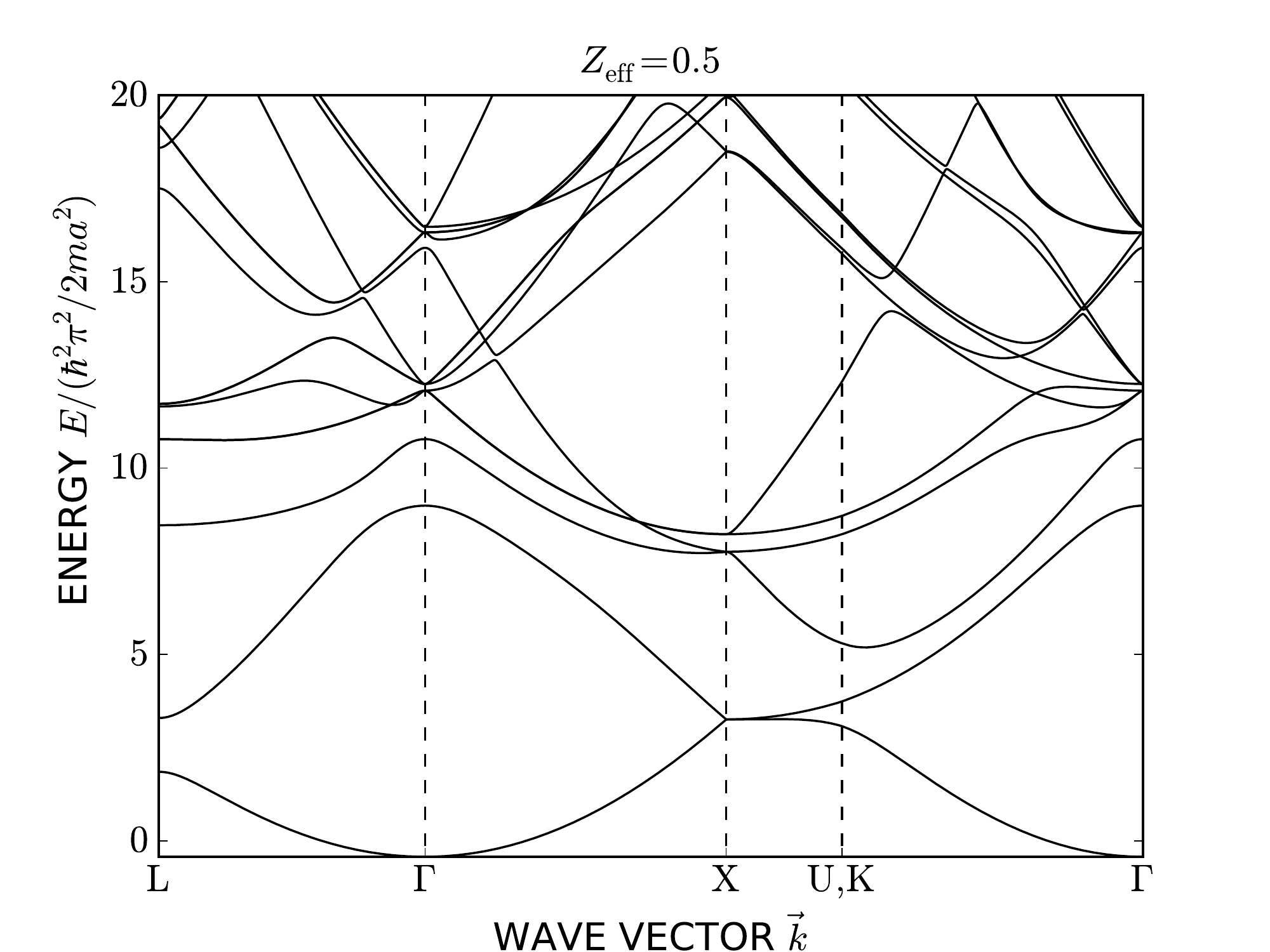}}%
\subfigure{\includegraphics[scale=0.4,clip=true,angle=0]{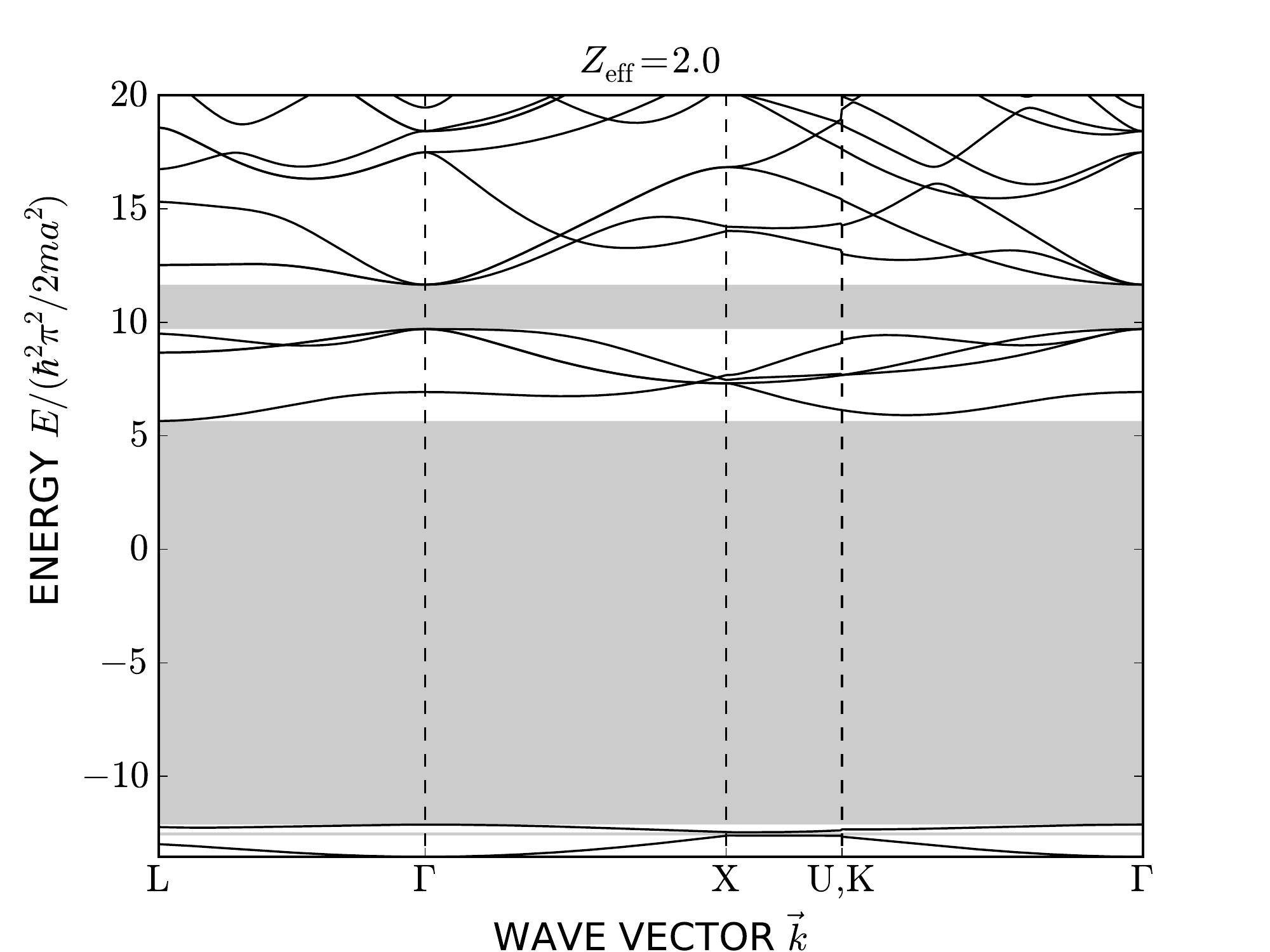}}%
\caption{Level splitting and band gap formation in a diamond cubic crystal as the strength of the Coulombic potential is
varied upwards from zero. From top left to bottom right, $Z_\text{eff} = 0,0.25,0.5,2.0$.  Energy 
gaps are shown in gray.  Letters along the horizontal axis represent points in the first Brillouin 
zone (Figure \ref{bz}).}
\label{bs}
\end{figure*}

\begin{figure*}[ht]
\vspace{-5pt}
\subtable[Selected matrix elements $\hat V_{k,k+G}$ for a non-Coulombic potential, where $|G|=n\pi/a$.  Units are electron
volts.  Remaining matrix elements are unchanged.]{
\begin{minipage}[t]{105pt}\vspace{40pt}
\begin{tabular}{| c || c | c |} \hline $n^2$ & $\hat V_{k,k+G}$ \\ \hline
%12 & -0.40 \\ 32 & -0.25 \\ 44 & \phantom{-}0.50 \\ 64 & \phantom{-}0.55 \\ 76 & \phantom{-}0.00 \\ 
0 &  -9.50 \\
12 &  \phantom{-}2.42 \\
32 &  \phantom{-}0.80 \\
44 &  -0.82 \\
64 &  \phantom{-}0.88 \\
76 &  \phantom{-}0.00 \\
\hline \end{tabular}
\vspace{46.3pt}\end{minipage}
\label{bs2:fc}}
\subfigure[Computed diamond cubic crystal band structure using the non-Coulombic potential described in \subref{bs2:fc}.]{
\begin{minipage}[t]{170pt}\vspace{0pt}
\hspace*{-7pt}\includegraphics[scale=0.4,clip=true,angle=0]{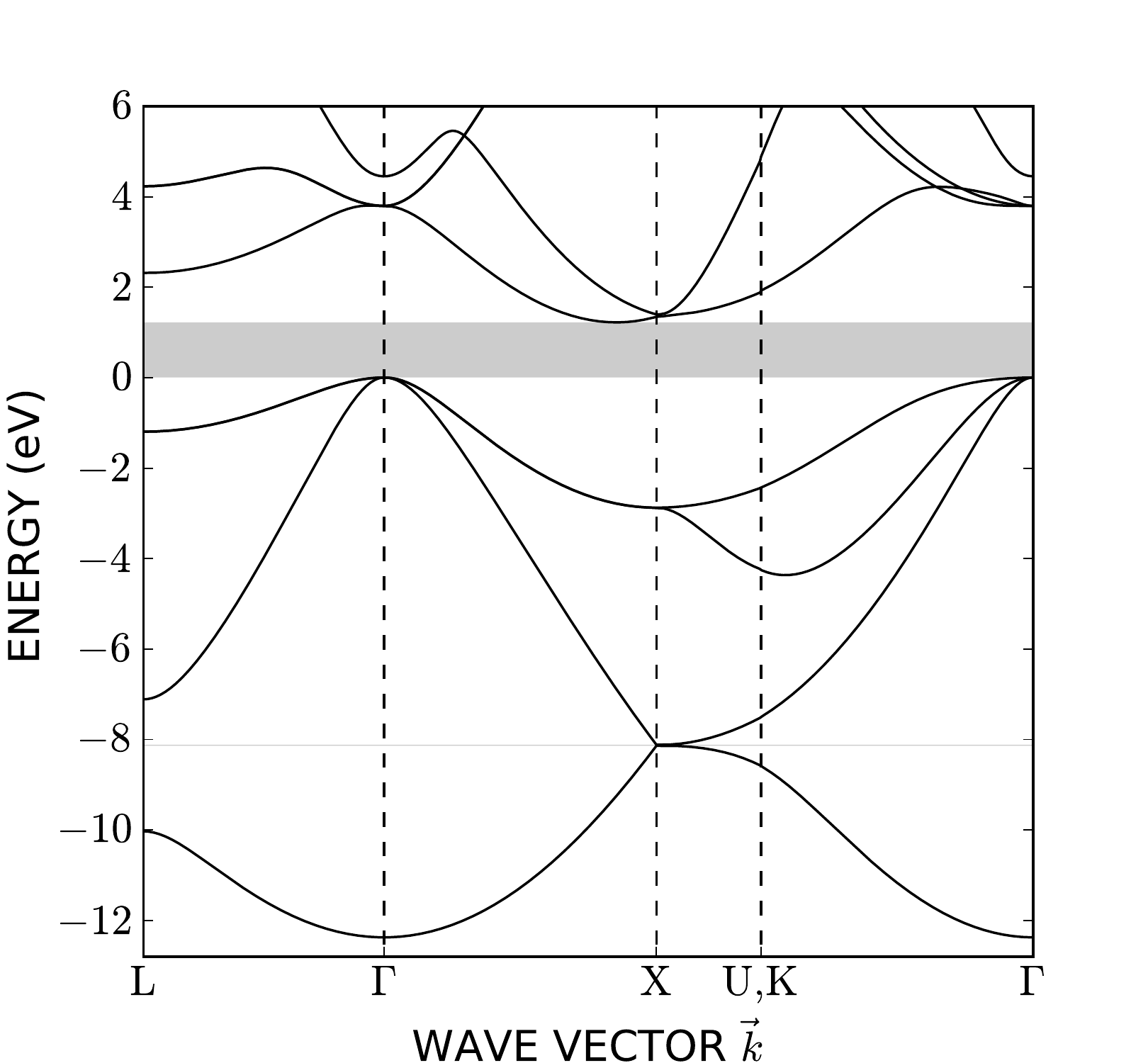}\vspace{3.9pt}
\end{minipage}
}
\subfigure[Band structure for Si.  Solid and dashed lines correspond to different techniques for incorporating e$^-$--e$^-$ interations.  Reproduced with permission from \cite{cohen1976}.  Copyright 1976 by The American Physical Society.]{
\begin{minipage}[t]{170pt}\vspace{15.5pt}
\includegraphics[scale=0.124,clip=true,angle=0]{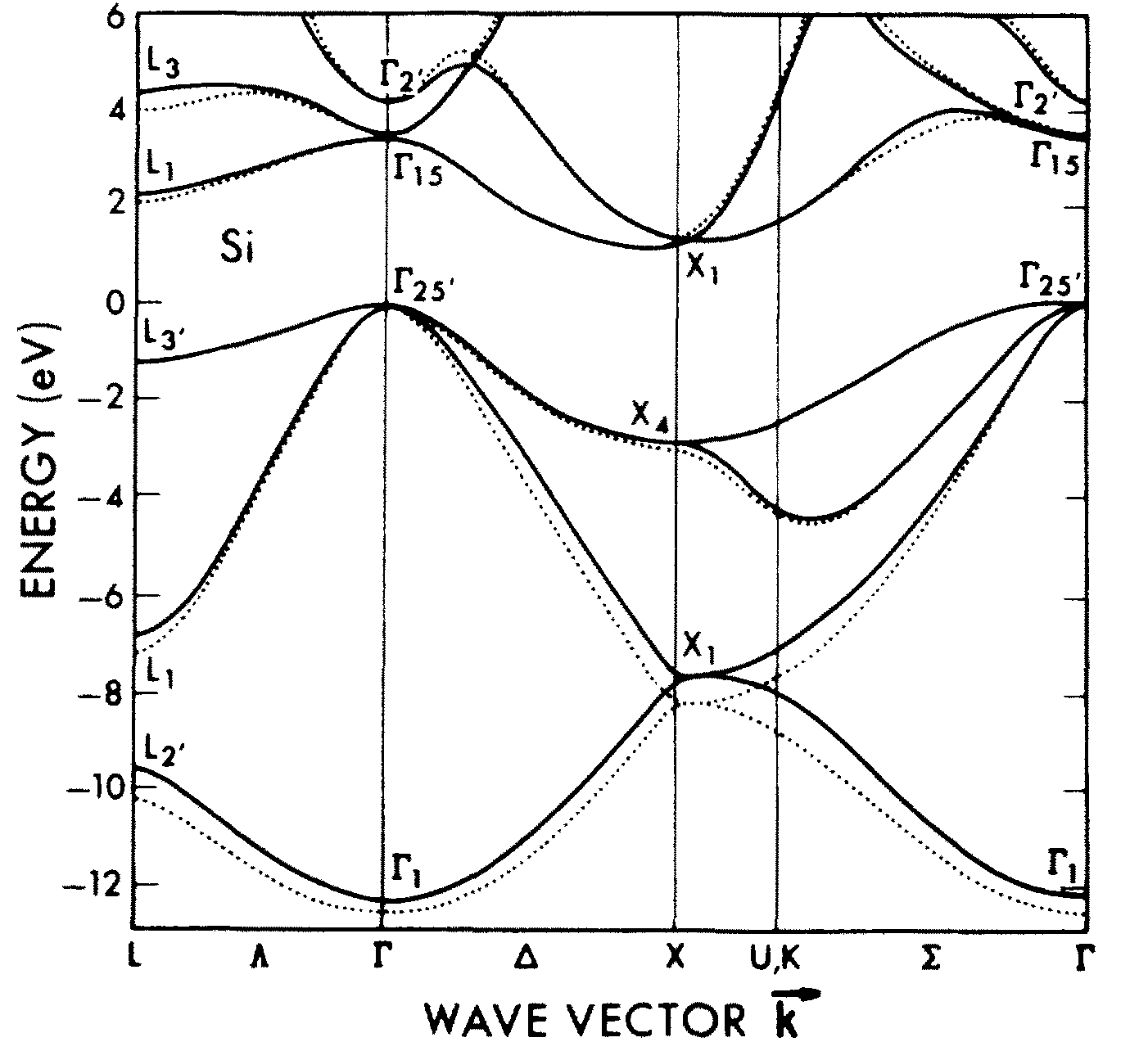}\vspace{2pt}
\end{minipage}
}
\caption{By modifying the potential, good agreement can be obtained with the band structure for Si shown in Figure 1 of \cite{cohen1976}.}
\label{bs2}
\end{figure*}

\noindent
Computed band structure diagrams for the diamond cubic crystal (Coulombic potential) are presented in Figure~\ref{bs} for four different values of $Z_\text{eff}$.  
The labels along the horizontal axis correspond to $\vec k$-space points in the first Brillouin zone (Figure \ref{bz}).  Note especially
that
\begin{itemize}
\item For a free electron, we expect $E=p^2/2m\propto k^2$.  In fact, when $Z_\text{eff}=0$, we find parabolic dispersion around the point $\vec k = \Gamma = 0$.
\item Along the path $L$--$\Gamma$, we recover much the same ``folded parabola'' structure we saw in one dimension, with branches of the
band diagram reflecting off the wall of the Brillouin zone (in this case, at $L$).
\item For small $Z_\text{eff}$, the effects of the perturbation should be most visible near crossings and degenerate branches of the band
diagram, where level repulsion is strongest.  This effect is observed as expected.  Note that even though the Coulomb potential gives
nonzero Fourier components for every nonzero $\vec k$, not all degeneracies are broken.  Part of the reason is due to zeros imposed by the
two-atom-per-FCC-site diamond cubic structure.
\item For a tightly bound electron (high $Z_\text{eff}$), we expect to find more or less isolated bands corresponding to the
orbitals of isolated atoms.  Since band spreading corresponds to tunneling in the tight binding picture, these bands describe
localized states.
\end{itemize}
In Figure~\ref{bs2}, we show how this simple model can incorporate the more complicated situation of many-particle interactions in an ad-hoc
way.  By selecting a few matrix elements essentially by hand, a very close correspondence can be obtained with results from more advanced
literature \cite{cohen1976}.  In the literature, these matrix elements would be calculated, if not from first principles, then from fitting
with experimental data such as x-ray reflectivity curves.


\section{Discussion}
As we have seen, it is possible to get a lot of mileage out of the
single-particle system.  By adopting more sophisticated approaches to
the potential due to e$^-$--e$^-$ interactions, \cite{cohen1976} and many others
have obtained models for band structure in real semiconductors with good
accuracy and substantial predictive power.  The single-particle framework we
have developed is simplistic insofar as it ignores the effects of spin-orbit
coupling, electron correlation, strains, electric and magnetic fields,
impurities, and finite-temperature effects.  It is nonetheless powerful enough
to make contact with the more advanced literature when supplied with an ad-hoc
choice of potential, and it remains beautiful in its simplicity.

\vspace{12pt}
\noindent
\subsection*{Acknowledgements}
The author is grateful to Nan Gu for conversations on finding matrix
elements in the plane wave basis, and to Yan Zhu and Francesco D'Eramo
for review and advice.
\vspace{70pt}
\nocite{*}
\bibliography{paper}
\end{document}